\documentclass[letterpaper, 10pt, conference]{IEEEtran}
\IEEEoverridecommandlockouts
\usepackage{cite}
\usepackage{amsmath,amssymb,amsfonts}
\usepackage{algorithmic}
\usepackage{graphicx}
\usepackage{textcomp}
\usepackage{xcolor}
\usepackage{xcolor,colortbl}
\usepackage{threeparttable}
\usepackage{stfloats}
\usepackage{booktabs}
\usepackage{makecell}
\usepackage{array}
\usepackage{multirow}

\definecolor{Gray}{gray}{0.85}
\def\BibTeX{{\rm B\kern-.05em{\sc i\kern-.025em b}\kern-.08em
    T\kern-.1667em\lower.7ex\hbox{E}\kern-.125emX}}
\begin{document}

\title{Clock-Shift-Aided Blind Doppler Matching in Passive SDR Monitoring Data for LEO Doppler Positioning
\thanks{This work was supported in part by the National Research Foundation of Korea (NRF), funded by the Ministry of Science and ICT (MSIT), Republic of Korea, under Grant RS-2024-00358298;
in part by the Unmanned Vehicles Core Technology Research and Development Program through the NRF and the Unmanned Vehicle Advanced Research Center (UVARC), funded by MSIT, Republic of Korea, under Grant RS-2020-NR046546;
in part by MSIT, Republic of Korea, under the Information Technology Research Center (ITRC) support program, supervised by the Institute of Information \& Communications Technology Planning \& Evaluation (IITP), under Grant IITP-2024-RS-2024-00437494;
in part by the Basic Science Research Program through the NRF, funded by the Ministry of Education, Republic of Korea, under Grant RS-2025-25425504; 
in part by Grant RS-2024-00407003 from the ``Development of Advanced Technology for Terrestrial Radionavigation System'' project, funded by the Ministry of Oceans and Fisheries, Republic of Korea;
and in part by the Korea Aerospace Administration (KASA), under Grant RS-2022-NR067078.}
}

\makeatletter
\newcommand{\linebreakand}{%
  \end{@IEEEauthorhalign}
  \hfill\mbox{}\par
  \mbox{}\hfill\begin{@IEEEauthorhalign}
}
\makeatother

\author{\IEEEauthorblockN{Seunghyeon Park}
\IEEEauthorblockA{
\textit{School of Integrated Technology} \\
\textit{Yonsei University}\\
Incheon, Korea \\
seunghyeon.park@yonsei.ac.kr}
\and
\IEEEauthorblockN{Gaeun Kim}
\IEEEauthorblockA{
\textit{School of Integrated Technology} \\
\textit{Yonsei University}\\
Incheon, Korea \\
kge1950@yonsei.ac.kr}
\and
\IEEEauthorblockN{Jiwon Seo${}^{*}$}
\IEEEauthorblockA{
\textit{School of Integrated Technology} \\
\textit{Yonsei University}\\
Incheon, Korea \\
jiwon.seo@yonsei.ac.kr}
{\small${}^{*}$ Corresponding author}
}

\maketitle
\begin{abstract}
Low Earth orbit (LEO) satellite signals have recently attracted increasing attention as opportunistic signals for positioning, navigation, and timing.
Most Doppler-based LEO positioning studies rely on tracked satellite signals or raw in-phase/quadrature (IQ) samples that preserve signal-level structures for acquisition, Doppler tracking, or satellite association.
In contrast, passive software-defined radio (SDR) monitoring systems often retain only time-frequency power scans, in which LEO Doppler traces appear as unlabeled ridges without raw IQ samples or identifiable signal structures.
This paper proposes a clock-shift-aided segment-based blind Doppler matching method that associates unlabeled Doppler traces with candidate satellites predicted using two-line element (TLE) data.
The proposed framework consists of slope-consistent Doppler segment generation, clock-shift-aided temporal alignment, and multi-metric satellite matching based on Doppler-rate and trajectory-shape consistency.
The proposed method was evaluated using a three-hour subset of open Starlink/OneWeb monitoring data collected in the United Kingdom.
The clock-offset search selected an effective offset of +19~s, for which 56 of 69 segments had at least one slope-compatible candidate, compared with 19 segments at zero shift.
Using the resulting satellite associations, the batch positioning error over the full three-hour period was 104.2~m, whereas the RMSE-based and slope-based simple-segment baselines failed to converge near the reference receiver location.
\end{abstract}


\section{Introduction}

Complementary navigation signals, including signals of opportunity, have been investigated to support positioning, navigation, and timing in Global Navigation Satellite System (GNSS)-challenged  environments \cite{Park25:Toward,Kim25:Enhancing,Kang25:Enhancing,Kim25:Set,Jeong24:Quantum,Kim26:Zonotope}.
Among them, low Earth orbit (LEO) satellite signals have recently attracted increasing attention as opportunistic signals for positioning, navigation, and timing (PNT) \cite{Prol22:Position,Stock25:Survey}.
Compared with medium Earth orbit GNSS satellites, LEO satellites are closer to terrestrial receivers and move with higher orbital velocities \cite{Reid18:Broadband}.
As a result, their received signals exhibit large and rapidly varying Doppler shifts, which can provide useful geometric information for receiver positioning when satellite orbit information and the carrier frequency are available \cite{Psiaki21:Navigation,Guo23:Instantaneous,Baron24:Implementation}.

Existing LEO Doppler positioning studies commonly rely on Doppler measurements extracted from tracked satellite signals or raw in-phase/quadrature (IQ) samples.
In single-satellite approaches, Doppler measurements are accumulated over time from an acquired or tracked satellite and used for batch or sequential positioning \cite{Psiaki21:Navigation,Neinavaie22:Acquisition}.
Because the satellite identity is already known in this setting, Doppler-to-satellite matching is not the main issue; instead, a long observation interval is often required to obtain sufficient Doppler geometry.

Multi-satellite approaches can improve the Doppler geometry by jointly using signals from multiple LEO satellites \cite{Kozhaya23:Multi,Allahvirdi25:Doppler,Hayek23:Assessment}.
In this case, however, each observed Doppler trace must be correctly associated with the satellite that generated it.
An incorrect association introduces an incorrect ephemeris and Doppler model into the positioning estimator, which can lead to large positioning errors or divergence.
Therefore, Doppler-to-satellite matching is a key prerequisite for multi-satellite LEO Doppler positioning.

Previous studies have addressed Doppler extraction, tracking, and satellite association under data conditions in which raw IQ samples or identifiable signal-level structures are available \cite{Stock25:Survey}.
For example, Starlink synchronization, carrier-phase, or burst structures have been exploited for acquisition and Doppler estimation \cite{Humphreys23:Signal,Neinavaie21:Exploiting,Stock24:Correlation,Khalife22:First}, and multi-constellation LEO beacons have been used for blind beacon estimation and Doppler association \cite{Kozhaya23:Multi}.
These approaches are effective when signal-level structures are preserved, but they are difficult to apply when the receiver stores only processed spectrum-level monitoring data.

This paper considers passive software-defined radio (SDR) monitoring data stored as time-frequency power scans rather than raw IQ samples.
Such data are commonly collected by low-cost monitoring systems for long-duration spectrum observation, spectrum occupancy analysis, interference monitoring, and RF sensing \cite{Rajendran18:Electrosense, CalvoPalomino20:Electrosense, Helbet21:Low}.
Because continuous IQ recording requires large storage capacity, practical monitoring systems often store frequency-domain power spectra instead of raw signal samples or navigation-ready observables.
Although this format is not originally intended for navigation, LEO Doppler traces can still appear as ridges in the time-frequency power representation.

The key challenge is that these Doppler ridges are unlabeled.
Multiple LEO satellites may pass through the monitored band during the same time interval, but the data do not contain raw IQ samples, beacon structures, or burst-level patterns typically used for conventional acquisition and tracking.
In addition, the observed frequency may include receiver-dependent frequency bias, and the logging time may include a coarse clock offset relative to the time system used for satellite orbit propagation \cite{Park26:LNSS}.
Therefore, to use passive SDR monitoring data as positioning observables, the unlabeled Doppler traces must first be reliably matched to their corresponding satellites.

This paper proposes a clock-shift-aided segment-based blind Doppler matching method for passive SDR monitoring data.
Here, the term blind refers to the unknown satellite identity of each observed Doppler trace.
The goal is not to outperform dedicated raw-IQ-based LEO positioning systems, but to demonstrate that non-positioning-oriented time-frequency power scans can be converted into usable Doppler positioning observables.

The proposed method groups spectral peaks into slope-consistent Doppler segments, estimates an effective clock offset between the monitoring data and two-line element (TLE)-based Doppler predictions, and matches each clock-shifted segment to a candidate satellite using a multi-metric cost function.
The method is evaluated using three hours of open Starlink/OneWeb monitoring data collected by a passive SDR receiver.
Using the matched Doppler measurements, the proposed method achieves a full three-hour batch positioning error of 104.2~m, whereas the RMSE-based and slope-based simple-segment baselines fail to converge near the reference receiver location.
These results show that passive SDR monitoring data can provide useful Doppler positioning information when unlabeled Doppler traces are properly segmented, temporally aligned, and matched to their corresponding satellites.

\section{Method}
\subsection{Algorithm Overview}

The proposed method performs blind Doppler matching using passive SDR monitoring data stored as time-frequency power scans.
As shown in Fig.~\ref{fig:overview_doppler}, multiple LEO satellites are observed by a passive SDR receiver, and their Doppler traces appear as unlabeled spectral peaks in the time-frequency domain.
Here, blind matching refers to the task of assigning each unlabeled Doppler segment to the satellite that generated it.

\begin{figure}[t]
\centering
\includegraphics[width=\linewidth]{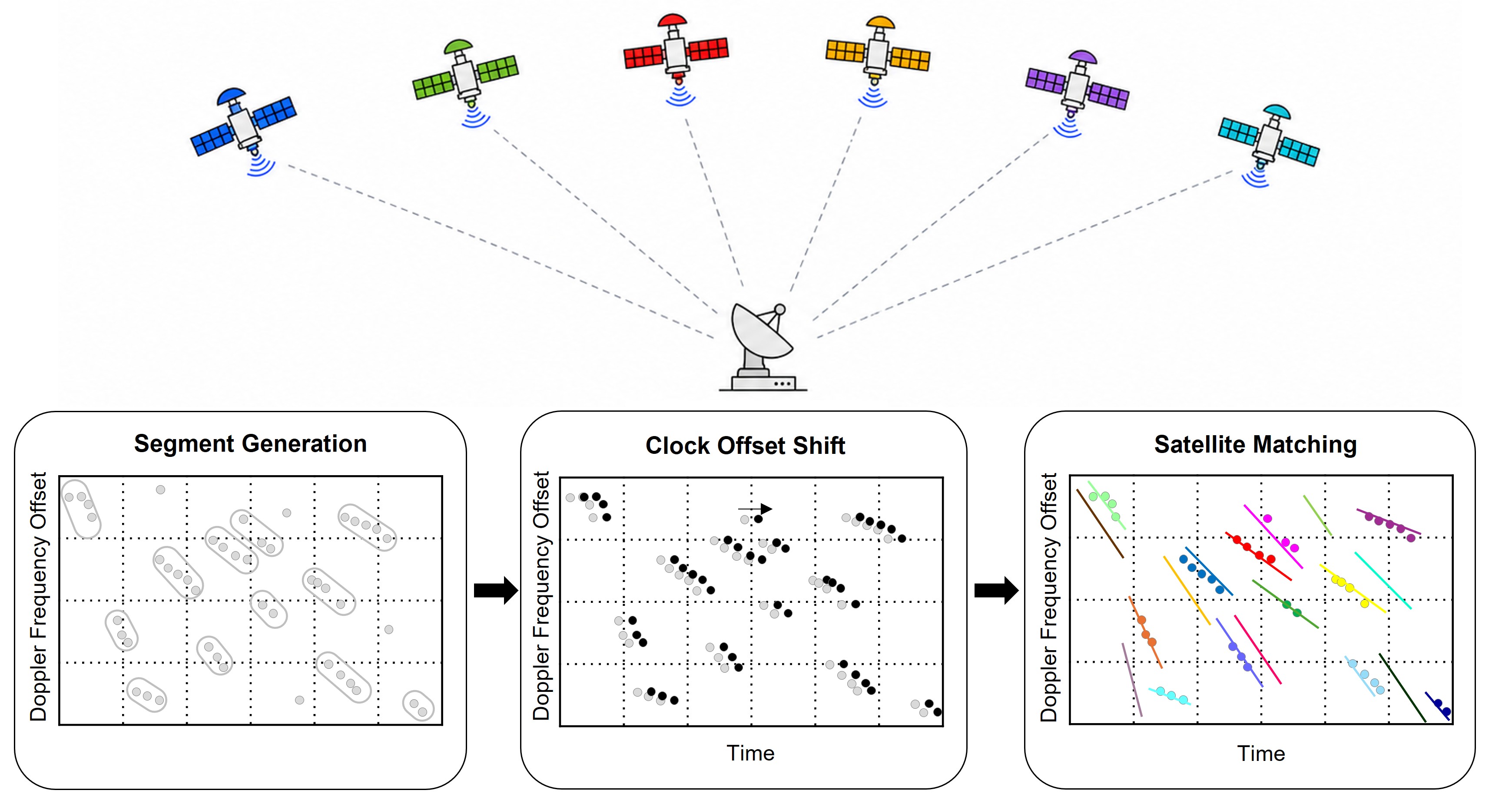}
\caption{Overview of the proposed blind Doppler matching framework for passive SDR monitoring data.}
\label{fig:overview_doppler}
\end{figure}

The proposed framework consists of three components: slope-consistent Doppler segment generation, clock-shift-aided temporal alignment, and multi-metric satellite matching.
The resulting satellite associations are then used for batch receiver positioning.

\subsection{Slope-Consistent Doppler Segment Generation}
\label{subsection:segment_generation}

The passive SDR monitoring data are first converted into spectral peaks by applying a threshold above the noise floor.
Because an isolated peak provides only an instantaneous frequency observation, adjacent peaks are grouped into short Doppler segments that represent local time-frequency ridges \cite{Iatsenko16:Extraction,Carmona99:Multiridge}.
If this grouping relies only on temporal and frequency proximity, peaks from different LEO Doppler traces can be incorrectly merged when multiple satellites appear in the same monitored band.
To reduce this risk, the proposed method additionally enforces local Doppler slope consistency as a gating criterion.

Let a detected peak be denoted by $(t_k,f_k,p_k)$, where $t_k$, $f_k$, and $p_k$ denote the time, frequency-bin location, and power of peak $k$, respectively.
The local peak-to-peak slope between a candidate peak and the previous peak in the current segment is defined as
\begin{equation}
\hat{a}_{k}^{\mathrm{loc}} = \frac{f_k - f_{k-1}}{t_k - t_{k-1}}.
\label{eq:local_slope}
\end{equation}
A candidate peak is appended to the current segment only when the time gap, frequency gap, and difference between $\hat{a}_{k}^{\mathrm{loc}}$ and the current segment slope are below predefined thresholds.
The updated segment is also checked using the RMSE and maximum absolute residual of a first-order fit.
These gating conditions are used to form short and reliable Doppler segments for subsequent satellite matching.

In the implementation, the maximum time gap and frequency gap were set to 15~s and 50~kHz, respectively.
The slope consistency threshold was set to 3500~Hz/s.
The segment fit RMSE threshold and maximum absolute residual threshold were set to 8000~Hz and 15000~Hz, respectively.
Segments shorter than 5~s were discarded.

\subsection{Clock Offset Correction}
\label{subsection:clock_offset}

The passive monitoring data were recorded over a long observation period using the receiver clock.
Thus, the recorded timestamps may not be exactly aligned with the UTC time axis used for TLE/SGP4 orbit propagation, which can cause incorrect satellite association for rapidly varying LEO Doppler trajectories \cite{Park24:CSAC,Vallado06:Revisiting,Ali98:Doppler}.
In this study, the estimated clock offset is used as an effective temporal alignment parameter for satellite association, not as a calibrated receiver clock error.
For each candidate offset $\tau$, candidate satellites are selected based on geometric visibility from the monitoring site at the shifted segment midpoint.
The offset that produces the largest number of slope-compatible segments is selected for subsequent matching.

Let $a_i^{\mathrm{obs}}$ denote the first-order fitted slope of observed segment $i$, where $i$ is the segment index, and let $a_j^{\mathrm{TLE}}$ denote the local TLE-based Doppler slope of candidate satellite $j$ for the shifted segment.
For each candidate offset $\tau$, $\mathcal{C}_i(\tau)$ denotes the set of candidate satellites that are geometrically visible from the monitoring site at the shifted midpoint of segment $i$.
The number of slope-consistent matches is computed as
\begin{equation}
N_{\mathrm{match}}(\tau)
=
\sum_{i \in \mathcal{S}}
\left[
\min_{j \in \mathcal{C}_i(\tau)}
\left|
a_i^{\mathrm{obs}} - a_j^{\mathrm{TLE}}
\right|
< \gamma_{\mathrm{shift}}
\right],
\label{eq:clock_match_count}
\end{equation}
where $\mathcal{S}$ is the set of observed segments, the bracketed condition equals 1 when it is true and 0 otherwise, and $\gamma_{\mathrm{shift}}$ is the slope agreement threshold.
The clock offset is then selected as
\begin{equation}
\tau^{*}
=
\operatorname*{arg\,max}_{\tau \in \mathcal{T}}
N_{\mathrm{match}}(\tau),
\label{eq:clock_offset}
\end{equation}
where $\mathcal{T}$ is the discrete set of candidate clock offsets.
In this study, $\mathcal{T}$ was searched from \textminus120~s to +120~s with a 1-s step, and $\gamma_{\mathrm{shift}}$ was set to 150~Hz/s.

After estimating $\tau^{*}$, the observed Doppler segments are shifted only for satellite matching.
The estimated offset is therefore used to evaluate the consistency between the observed Doppler segments and the TLE/SGP4-predicted candidate trajectories on a common time axis.
The final batch positioning stage uses the satellite associations obtained after this matching process.

\subsection{Multi-Metric Doppler Matching}
\label{subsection:multi_metric_matching}

After clock offset correction, each observed segment is compared with TLE/SGP4-predicted candidate satellite trajectories generated using a coarse receiver position \cite{Vallado06:Revisiting}.
A city-level receiver position with an approximately 3-km error is assumed, and the detected peak frequencies are converted to a Doppler-like frequency axis by subtracting a common reference frequency.

Let $f_{i,k}^{\mathrm{obs}}$ denote the observed Doppler-like sample $k$ in segment $i$ at time $t_{i,k}$, where $k=1,\ldots,N_i$.
The shifted time coordinate is
\begin{equation}
\tilde{t}_{i,k}=t_{i,k}+\tau^{*}.
\label{eq:shifted_observed_time}
\end{equation}
For candidate satellite $j$, the theoretical Doppler frequency is computed from the TLE-based range rate as \cite{Psiaki21:Navigation}
\begin{equation}
f_{j}^{\mathrm{TLE}}(t) = - \frac{f_c}{c} \dot{\rho}_{j}(t),
\label{eq:theory_doppler}
\end{equation}
where $f_c$ is the carrier frequency, $c$ is the speed of light, and $\dot{\rho}_{j}(t)$ is the range rate between the coarse receiver position and satellite $j$.

The proposed method combines Doppler-rate and trajectory-shape consistency, which have been commonly used for LEO Doppler tracking and association \cite{Neinavaie22:Acquisition, Stock24:Correlation, Kozhaya25:Unveiling}.
The first metric is the Doppler slope difference:
\begin{equation}
\Delta a_{i,j}
=
\left| a_i^{\mathrm{obs}} - a_j^{\mathrm{TLE}} \right|.
\label{eq:slope_difference}
\end{equation}

The second metric is the residual RMSE after removing a candidate-wise constant frequency offset:
\begin{equation}
\mathrm{RMSE}_{i,j}
=
\sqrt{
\frac{1}{N_i}
\sum_{k=1}^{N_i}
\left(
f_{i,k}^{\mathrm{obs}}
-
f_j^{\mathrm{TLE}}(\tilde{t}_{i,k})
-
\delta_{i,j}
\right)^2
},
\label{eq:residual_rmse}
\end{equation}
where $\delta_{i,j}$ is the mean residual
$f_{i,k}^{\mathrm{obs}}-f_j^{\mathrm{TLE}}(\tilde{t}_{i,k})$
over segment $i$, making the metric insensitive to a constant frequency bias.

Using $A_0=500$~Hz/s and $F_0=2000$~Hz, the two metrics are normalized, bounded above by one, and combined as
\begin{equation}
J_{i,j}
=
\frac{\Delta a_{i,j}}{A_0}
+
\frac{\mathrm{RMSE}_{i,j}}{F_0},
\label{eq:matching_cost}
\end{equation}
where each normalized term is bounded above by one before summation.
Nonuniform weighting was also tested, and varying the slope-term weight from 0.2 to 0.8 with complementary residual weights produced identical satellite associations and the same final positioning error; therefore, equal weighting was adopted.
The final matched satellite is selected as
\begin{equation}
j_i^{*}
=
\operatorname*{arg\,min}_{j \in \mathcal{C}_i(\tau^{*})}
J_{i,j}.
\label{eq:final_candidate}
\end{equation}

\section{Experiment}

\subsection{Dataset}

The proposed method was evaluated using a three-hour subset of an open Starlink/OneWeb monitoring dataset collected in Altrincham, United Kingdom, in March 2023 \cite{Watson23:Raw}.
The original dataset was collected using an RTL-SDR-based receiver for satellite constellation monitoring rather than positioning \cite{Watson23:Raw, Watson24:Starlink}.
The analyzed subset covers the period from 00:00 to 03:00 UTC on March 11, 2023.

The dataset is stored as time-frequency power scans rather than raw IQ samples; each row corresponds to a spectrum scan, and the columns contain frequency-bin power values over the monitored band.
Thus, the input to the proposed method is a spectrogram-like power matrix rather than a demodulated beacon signal or a navigation observable.
The nominal reference frequency used for Doppler conversion was 11.325~GHz, and Starlink and OneWeb TLEs for the analyzed period were used to generate candidate satellite trajectories and for visibility-based candidate filtering.

\subsection{Comparison Methods}

To the best of the authors' knowledge, blind Doppler matching methods for passive SDR monitoring data without raw IQ samples or identifiable signal-level structures have not yet been well established.
Therefore, two simple baseline methods were designed using commonly used Doppler similarity criteria in LEO Doppler tracking and matching studies \cite{Kozhaya25:Unveiling,Stock24:Correlation}.
For the baseline methods, Doppler segments were first generated using only temporal and frequency continuity, without slope-consistency gating.
Three matching methods were compared:
\begin{itemize}
\item \textbf{Case A: Proposed method.}
\item \textbf{Case B: Simple-segment RMSE-based matching.}
Each segment is matched to the candidate satellite that provides the smallest residual RMSE.
\item \textbf{Case C: Simple-segment slope-based matching.}
Each segment is matched to the candidate satellite that provides the smallest Doppler slope difference.
\end{itemize}

Cases B and C are not intended as component-wise ablations.
Instead, they serve as simple end-to-end baselines to examine whether commonly used single-metric Doppler similarity criteria, combined with simple time-frequency segment generation, are sufficient for positioning from passive SDR monitoring data.

\subsection{Evaluation Metrics}

The evaluation primarily focuses on whether the satellite associations obtained by each matching method can be used for receiver positioning.
The positioning performance was evaluated using the 3D Earth-centered, Earth-fixed (ECEF) error relative to the reference receiver position.

Because the monitoring data were collected over a narrow frequency span, only a small number of Doppler peaks were available at each epoch.
In many epochs, only one or two visible Doppler traces appeared in the monitored band, making single-epoch positioning infeasible.
Therefore, only batch positioning was performed in this study.

For an accepted Doppler sample $n$ associated with satellite $s_n$, the Doppler observation is modeled as
\begin{equation}
z_n
=
-\frac{f_c}{c}
\dot{\rho}_{s_n}(\mathbf{r},t_n)
+
\epsilon_n,
\label{eq:positioning_measurement_model}
\end{equation}
where $z_n$ is the Doppler-like observation after removal of the candidate-wise constant frequency offset, $\mathbf{r}$ is the stationary receiver ECEF position, and $\epsilon_n$ is the residual Doppler error.
The residual term $\epsilon_n$ also includes unmodeled receiver- and satellite-side frequency errors, including oscillator drift and transmitter frequency variations, which are not separately estimated in the present formulation.
The batch solution estimates the receiver position $\mathbf{r}$ by minimizing the squared Doppler residuals over the accepted samples using a bounded trust region reflective algorithm for nonlinear least squares.
The nonlinear solver was initialized with the initial ECEF position estimate $\mathbf{r}_0=[0,0,0]^T$~m; this value was not used as a receiver-position prior.

In addition to positioning error, the Doppler residual RMSE after positioning was computed as a supplementary metric.
This residual analysis was used to examine whether the matched Doppler measurements were dynamically consistent with the estimated receiver position and the matched satellite ephemerides.

\begin{figure*}[t]
\centering
\includegraphics[width=0.95\textwidth]{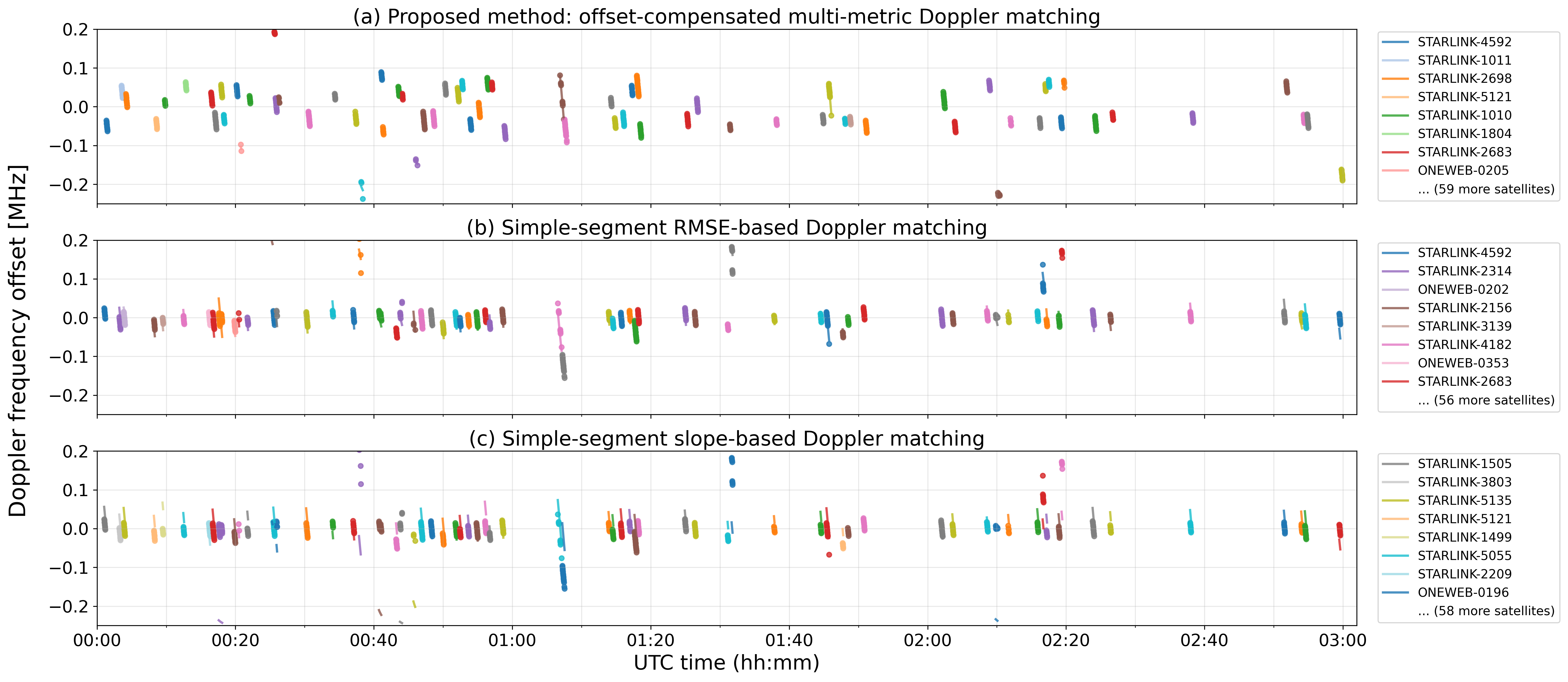}
\caption{Doppler matching results for the proposed method and the two simple-segment baseline methods. Solid curves represent TLE-based theoretical Doppler trajectories, and colored points represent observed spectral peaks assigned to the corresponding satellite trajectories.}
\label{fig:matching_result}
\end{figure*}

\section{Results}

\subsection{Doppler Matching Results}

The proposed blind Doppler matching method was applied to the real-world passive monitoring dataset.
Using the method described in Section~\ref{subsection:segment_generation}, 69 slope-consistent Doppler segments were generated from the extracted spectral peaks.

The temporal-alignment search was then performed using the method described in Section~\ref{subsection:clock_offset}.
At zero shift, 19 of the 69 slope-consistent segments had at least one slope-compatible candidate.
Maximizing the objective in \eqref{eq:clock_offset} yielded an effective temporal-alignment offset of +19~s, for which 56 segments had at least one slope-compatible candidate.
Because this offset was selected by maximizing $N_{\mathrm{match}}(\tau)$, the increase in compatible segments was interpreted as an internal matching diagnostic rather than an independent validation of positioning performance.
The selected offset was therefore used only to improve the temporal consistency of the satellite matching process.

After clock offset correction, final satellite matching was performed using the multi-metric cost.
Fig.~\ref{fig:matching_result} shows the Doppler matching results for the proposed method and the two baseline methods.
In the figure, the solid curves represent TLE-based theoretical Doppler trajectories, and the colored points represent observed spectral peaks assigned to the corresponding satellite trajectories.

For the proposed method, the observed peaks are generally assigned to satellite trajectories with similar Doppler trends over the three-hour observation period.
In contrast, the two baseline methods show several assignments that appear inconsistent with the local Doppler trends of the observed peaks.

\subsection{Positioning Results}

Using the satellite associations obtained from the Doppler matching process, Doppler-only cumulative batch positioning was performed.
As shown in Fig.~\ref{fig:batch_positioning_result} and Table~\ref{table:positioning_error}, the proposed method did not converge for the 20-min window, producing a 3D ECEF error of 1203.9~km.
For the evaluated record starting at 00:00 UTC, the error decreased to 619.8~m at 40~min and reached 104.2~m at 180~min as more Doppler observations were accumulated.
In contrast, the RMSE-based and slope-based baselines remained at 78.5~km and 5500.8~km, respectively, for the 180-min solution.
Although the final positioning error is not comparable to that of dedicated LEO navigation receivers, convergence to approximately 100~m suggests that passive SDR monitoring data can still support batch Doppler positioning when the Doppler traces are properly matched to satellites.
The relatively long convergence time mainly results from the limited instantaneous Doppler geometry, because many epochs contain only one or two visible Doppler traces.
Monitoring a wider frequency band to observe more simultaneous LEO signals could reduce the observation time required for positioning.

\begin{figure}[t]
\centering
\includegraphics[width=\linewidth]{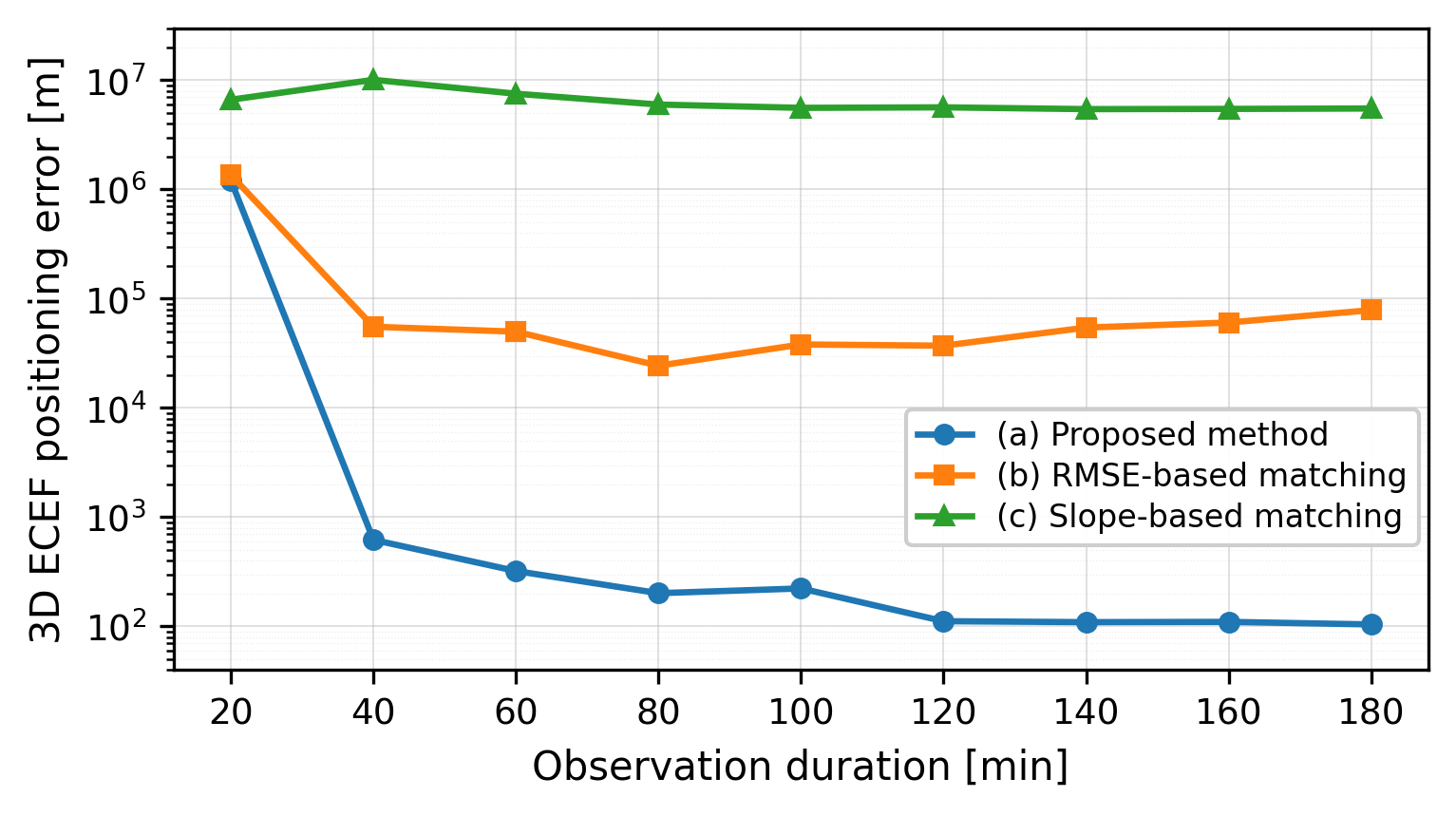}
\caption{Cumulative 3D ECEF batch positioning errors for the proposed method and the two baseline methods. The vertical axis is shown on a logarithmic scale to display the overall error distribution.}
\label{fig:batch_positioning_result}
\end{figure}

\begin{table}[t]
\centering
\caption{Cumulative 3D ECEF batch positioning errors for selected observation windows}
\renewcommand{\arraystretch}{1.5}
\begin{tabular}{l||ccc}\hline
Window & Proposed & RMSE-based & Slope-based \\
 \hline
20~min & 1203.9~km & 1354.6~km & 6621.0~km \\
40~min & 619.8~m & 55.3~km & 10069.2~km \\
60~min & 321.1~m & 49.9~km & 7512.3~km \\
80~min & 201.4~m & 24.3~km & 5978.6~km \\
120~min & 111.5~m & 37.1~km & 5637.3~km \\
180~min & 104.2~m & 78.5~km & 5500.8~km \\
 \hline
\end{tabular}
\label{table:positioning_error}
\end{table}

\subsection{Doppler Residual Analysis}
Table~\ref{table:doppler_residual} summarizes the Doppler residual RMSE values after the full three-hour batch positioning.
The proposed method achieved a residual RMSE of 1.59~kHz, whereas the RMSE-based and slope-based baselines yielded 17.09~kHz and 45.66~kHz, respectively.
Although the proposed method has a kilohertz-level residual, which is expected from the low-cost passive monitoring setup and spectrogram-based Doppler extraction, its substantially smaller residual indicates better consistency with the estimated receiver position and satellite dynamics than the residuals of the baselines.

\begin{table}[t]
\centering
\caption{Comparison of Doppler residual RMSE after batch positioning}
\renewcommand{\arraystretch}{1.5}
\begin{tabular}{l||ccc}\hline
 & Proposed & RMSE-based & Slope-based \\
 \hline
Residual RMSE & 1.59~kHz & 17.09~kHz & 45.66~kHz \\
 \hline
\end{tabular}
\label{table:doppler_residual}
\end{table}

\section{Conclusion}

This paper presented a clock-shift-aided segment-based blind Doppler matching method for passive SDR monitoring power scans.
The proposed method generates slope-consistent Doppler segments, estimates an effective temporal-alignment offset for satellite association, and matches each segment to a TLE-predicted candidate trajectory using multiple Doppler-shape metrics.
The purpose of this study was to demonstrate that non-positioning-oriented time-frequency power scans can be converted into usable Doppler positioning observables after reliable satellite association.

For one three-hour Starlink/OneWeb monitoring subset, maximizing the temporal-alignment objective yielded an offset of +19~s, for which 56 of 69 segments had at least one slope-compatible candidate, compared with 19 segments at zero shift.
Using the resulting associations, the 180-min cumulative batch solution achieved a 3D ECEF error of 104.2~m and a post-fit Doppler residual RMSE of 1.59~kHz, whereas the simple-segment baselines did not converge near the reference location.
Because the dataset does not provide independent timing ground truth or per-segment satellite labels, the estimated temporal offset and the resulting associations were evaluated through slope-compatibility diagnostics and end-to-end batch positioning consistency rather than direct clock-error or association-accuracy measurements.
Because the present evaluation is limited to a single three-hour measurement interval, further validation using additional observation periods and independent datasets is needed to assess the generalizability of the empirically selected parameters.
Future work will evaluate the method using labeled or raw-IQ datasets, multiple receiver priors, additional time periods, and explicit oscillator and transmitter-frequency models.

\section*{Acknowledgment}

The authors acknowledge the use of a generative AI tool (ChatGPT, OpenAI) to assist with grammar and language editing during manuscript preparation.  
The AI tool was not used to generate technical content, ideas, data, or citations.  
All technical content, methodology, analysis, and conclusions are the sole work of the authors.

\bibliographystyle{IEEEtran}
\bibliography{mybibfile, IUS_publications}

\end{document}